\documentclass[runningheads]{llncs}
\usepackage[T1]{fontenc}
\usepackage{graphicx,verbatim}
\usepackage{caption}

\usepackage{amssymb, amsmath, bm, latexsym,comment}
\usepackage{multirow}
\usepackage{xcolor}
\usepackage{subfigure}
\usepackage{indentfirst}
\usepackage{setspace}
\usepackage{verbatim}
\usepackage{array}
\usepackage{arydshln}
\usepackage[misc]{ifsym} 

\providecommand{\Leireftb}[1]{Table~\ref{#1}}

\providecommand{\citep}[1]{\cite{#1}}

\begin{document}

\title{Personalized 4D Whole Heart Geometry Reconstruction from Cine MRI for Cardiac Digital Twins}
\titlerunning{4D Whole Heart Reconstruction from Cine MRI}

\author{Xiaoyue Liu \inst{1} \and 
Xicheng Sheng \inst{2} \and 
Xiahai Zhuang \inst{2} \and 
Vicente Grau \inst{3} \and 
Mark YY Chan \inst{4,5} \and 
Ching-Hui Sia \inst{4,5} \and 
Lei Li\inst{1}${^{(\textrm{\Letter})}}$ } 
\authorrunning{Liu et al.}
\institute{Department of Biomedical Engineering, National University of Singapore, Singapore \and
School of Data Science, Fudan University, Shanghai, China \and
Department of Engineering Science, University of Oxford, Oxford, UK \and
Department of Medicine, National University of Singapore, Singapore \and
Department of Cardiology, National University Heart Centre Singapore, Singapore\\
\email{lei.li@nus.edu.sg}}


\maketitle  
\begin{abstract}

Cardiac digital twins (CDTs) provide personalized in-silico cardiac representations and hold great potential for precision medicine in cardiology.
However, whole-heart CDT models that simulate the full organ-scale electromechanics of all four heart chambers remain limited. 
In this work, we propose a weakly supervised learning model to reconstruct 4D (3D+t) heart mesh directly from multi-view 2D cardiac cine MRIs. 
This is achieved by learning a self-supervised mapping between cine MRIs and 4D cardiac meshes, enabling the generation of personalized heart models that closely correspond to input cine MRIs.  
The resulting 4D heart meshes can facilitate the automatic extraction of key cardiac variables, including ejection fraction and dynamic chamber volume changes with high temporal resolution. 
It demonstrates the feasibility of inferring personalized 4D heart models from cardiac MRIs, paving the way for an efficient CDT platform for precision medicine.  
The code will be publicly released once the manuscript is accepted.

\keywords{Cine MRI \and 4D Heart Reconstruction \and Whole Heart \and Self-Supervised Mapping \and 
 Cardiac Digital Twins.}

\end{abstract}

\section{Introduction}

Cardiovascular diseases remain the leading cause of mortality worldwide.
Cardiac digital twin (CDT) technology has emerged as a powerful approach for creating patient-specific virtual heart models, enabling real-time analysis of cardiac structure and function \citep{journal/PTRS/niederer2020}. 
By offering detailed insights into the underlying mechanisms of the heart, CDT has the potential to revolutionize cardiac diagnosis and treatment \citep{journal/EHJ/corral2020,journal/NC/arevalo2016}.
A typical CDT workflow consists of two key stages: anatomical twinning and functional twinning \citep{journal/MedIA/gillette2021,journal/TMI/li2024}.
Anatomical twinning involves extracting 3D heart geometry from images and identifying pathological regions when present. 
Considering the dynamic nature of cardiac motion, 4D (3D+t) geometry is typically required for a more comprehensive representation.
Cine MRI can be used for this purpose, as it provides non-invasive visualization of cardiac anatomy and motion throughout the cardiac cycle. 
However, cine MRI typically acquires sparse and intersecting 2D image planes, i.e., short-axis (SAX) and long-axis (LAX) slices, which limits spatial resolution and hinders the construction of a fully detailed 4D representation of the heart. 
These constraints can impact the accuracy of anatomical twinning and downstream functional twinning which involves the simulation of cardiac electromechanics.

Conventional cardiac geometry reconstruction frameworks generally consist of two steps, i.e., image segmentation, followed by mesh generation based on contours derived from the segmentation.  
This is mainly because direct volumetric reconstruction from cine MRIs is challenging due to the inherent sparsity and anisotropy of the data. 
By first segmenting the cardiac structures, the extracted contours can serve as geometric constraints to guide the mesh generation. 
However, cine MRI only provides a sparse representation of the actual 3D geometry of the human heart. 
Consequently, traditional iso-surfacing algorithms, such as marching cubes, struggle to generate smooth and anatomically accurate meshes due to the irregular spacing and insufficient volumetric information in the input data.
To solve this, many previous studies employed mesh adaptation approaches, such as template mesh deformation \cite{journal/JI/villard2018,journal/CMPB/hu2023}, statistical shape model (SSM) \cite{journal/PTRSA/banerjee2021}, and B-spline surface reconstruction \cite{journal/MRM/odille2018,journal/BMM/bennati2023}. 
Furthermore, image interpolation based methods have also been applied to reconstruct high-resolution 3D geometry \cite{journal/TMI/ukwatta2015}. 
Nonetheless, these techniques are labor-intensive, requiring complex manual initialization or adjustments of optimizer parameters, which significantly hinders their feasibility for real-time applications.

Recently, deep learning-based methods have achieved promising performance for efficient 3D cardiac geometry reconstruction.
Similarly, they generally rely on pre-generated segmentation and then directly convert sparse contour point clouds into 3D meshes via point completion network \cite{journal/MedIA/beetz2023} or graph convolution network (GCN) based template deformation \cite{journal/MedIA/chen2021,conf/ICCV/ye2023}. 
Instead of directly performing interpolation on images, Chang et al. \cite{conf/MICCAI/chang2022} developed a latent space-based generative method to simultaneously predict 2D SAX segmentation and 3D volume by interpolating latent codes. 
Biffi et al. \cite{conf/ISBI/biffi2019} first segmented cine MRI and then reconstructed a high-resolution 3D volume using a conditional variational autoencoder, incorporating features from one SAX and two LAX segmentation.
For 4D cardiac reconstruction, Yuan et al. \cite{conf/ICCV/yuan2023} decoupled cardiac motion and shape from the given sparse contour point cloud sequences based on the neural motion model and deep SSM model.
Recently, there are several deep learning based studies directly reconstruct meshes from cardiac image \cite{journal/MedIA/kong2021,journal/MedIA/laumer2023,journal/arxiv/chen2024}.
For example, Kong et al. \cite{journal/MedIA/kong2021} directly predicted whole heart surface meshes from volumetric CT and MRI data via GCN based pre-defined mesh template deformation.
Chen et al. \cite{journal/arxiv/chen2024} employed a deep marching tetrahedra model that discretized 3D space into a deformable tetrahedral grid, assigning each vertex a signed distance value for 4D biventricular reconstruction.
Laumer et al. \cite{journal/MedIA/laumer2023} reconstructed 4D whole heart mesh from 2D echocardiography video data via task-tailored autoencoder models.
However, their approach relied on single-view, single-slice 2D data, which provided limited spatial information of the heart and thus constrained the reconstruction accuracy.
In general, current work either relied on high-resolution volumetric images or solely reconstructed part of the whole heart.

In this work, we present a novel weakly supervised model to infer personalized whole-heart meshes from multi-view cine MRI.
Given unpaired cine MRI and 4D heart mesh, the model can efficiently map the cine MRI to a 4D heart mesh via the generative domain translation.
Specifically, we utilize the domain-specific autoencoder networks to extract the compact latent representations of both cine MRIs and the mesh videos. 
Then, the mapping between the image and heart mesh video latent spaces can be learned to ensure the generated shapes align with the cardiac deformation space. 
To the best of our knowledge, this is the first study to directly reconstruct a 4D whole-heart mesh from cine MRIs.

\begin{figure*}[t]\center
 \includegraphics[width=0.99\textwidth]{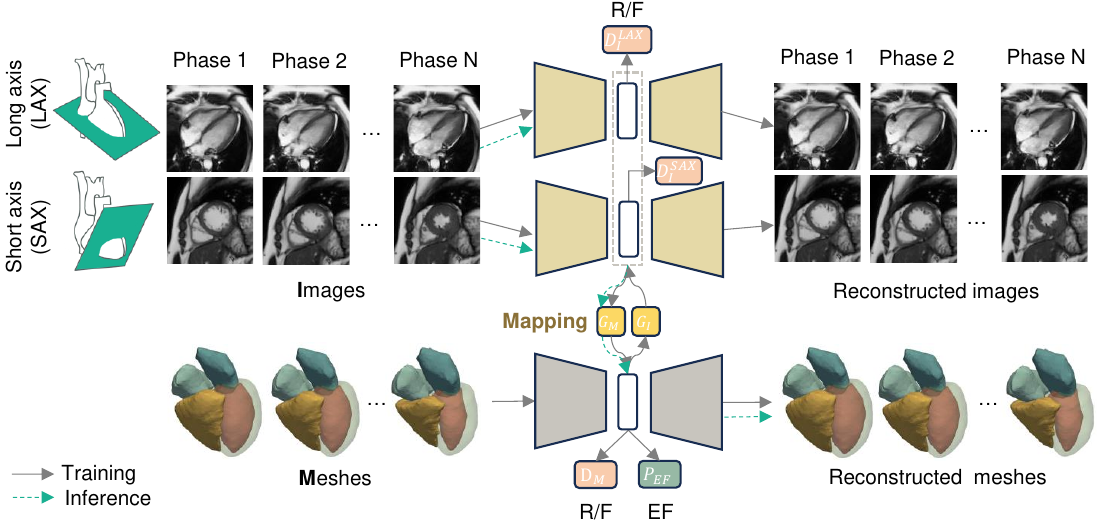}\\[-2ex]
   \caption{Illustration of the multi-view cine-MRI based 4D whole heart mesh reconstruction framework. Note that the diagram only shows a single short-axis (SAX) slice as an example, even though three representative slices in SAX view are selected to capture comprehensive spatial information. }
\label{fig:method:framework}
\end{figure*}

\section{Methodology}

Figure~\ref{fig:method:framework} provides an overview of the proposed 4D whole heart reconstruction model, consisting of domain-specific autoencoders (AE) and cycle-mapping modules.
Image and mesh AE networks are designed to extract the compact latent representations of cine MRIs and mesh videos, respectively (Sec.~\ref{method:AE}). 
To train the generator network mapping between the image and heart mesh video latent space, we optimize a cycle-consistency loss to enforce reversibility and an adversarial loss with two discriminators distinguishing real from fake representations (Sec.~\ref{method:mapping}). 
Additionally, ejection fraction predictor acts as a weak regularizer to guide generator training.
Finally, Sec.~\ref{method:inference} presents the details of the reconstruction model for the personalized inference of 4D whole heart mesh.

\subsection{Domain-Specific Autoencoder for Feature Extraction} \label{method:AE}

To extract the cardiac spatial and motion features from cine MRIs, we employ the image-specific AE (namely \(AE_I\)) on two different cine views, i.e., LAX and SAX views.
LAX view includes single slice, while SAX view is the stack of several slices but we only select three representative slices, i.e., apical, middle, and basal slices, for simplification.
Given a sequence of cine frames, $\{I_t^{SAX}, I_t^{LAX}\}_{t=1}^{N}$, where $t$ is the frame ID and $N$ is the number of frames in the cine data, we can extract the corresponding low-dimensional manifold \(\mathcal{Z}_I\) for each view and fuse them as cine imaging latent space.
To achieve this, we integrate prior knowledge of periodicity by representing the latent trajectory as a circular motion in the first two dimensions, with additional parameters controlling shape and dynamics. 
The \(AE_I\) encoder, consisting of a CNN and an LSTM, extracts features and predicts trajectory parameters \(\phi_I\) as its final state. 
Instead of directly using these parameters, \(AE_I\) decoder reconstructs observations based on the learned latent trajectory to ensure temporal consistency.
The optimization of \(\text{AE}_I\) is based on minimizing a regularized image reconstruction loss:
\begin{equation}
    \mathcal{L}^{recon}_{I} = \frac{1}{N} \sum_{t=1}^{N} \| I_t - \hat{I}_t \|^2 + \mathcal{R}(\phi_I),
\end{equation}
where \( I_jt \) and \( \hat{I}_t \) are the input and reconstructed cine frames, respectively, and \(\mathcal{R}(\phi_I)\) is the regularizer to constrain the latent trajectory parameters. 

Given a mesh video as a sequence of 3D heart meshes with corresponding time steps, represented as $\{M_t\}_{t=1}^{N}$, we can similarly extract their spatial and motion features via a mesh-specific AE (namely \(AE_M\)).
It consists of an encoder that encompasses a feature extractor network and an LSTM which outputs as its final state compressed representation \(\phi_M\).
The feature extractor and the \(AE_M\) decoder is based on graph neural networks.
This is because each surface mesh $M_t$ consists of vertices and faces forming polygonal surfaces and can be described as an undirected graph $G = (\mathcal{V}, \mathcal{E})$, where $\mathcal{V} = \{v_1, ..., v_n\}$ is the set of $n$ vertices, and $\mathcal{E} = \{(v_i, v_j)\}$ (for $i \neq j$) represents the set of edges connecting the vertices. 
The neighborhood of a vertex $v_i$, denoted as $\mathcal{N}_i$, is defined as the set of vertices $v_j$ such that $(v_i, v_j) \in \mathcal{E}$, i.e., $\mathcal{N}_i = \{v_j \mid (v_i, v_j) \in \mathcal{E}\}$.
The graph structure is represented by an adjacency matrix $A$ and vertex $v_i$ is associated with a feature vector.
The optimization of \(\text{AE}_M\) is achieved by minimizing a regularized mesh reconstruction loss:
\begin{equation}
    \mathcal{L}^{recon}_{M} = \frac{1}{N} \sum_{t=1}^{N} \| M_t - \hat{M}_t \|^2 + \mathcal{R}(\phi_M),
\end{equation}
where \( M_jt \) and \( \hat{M}_t \) are input and reconstructed mesh video, respectively, and \(\mathcal{R}(\phi_M)\) is the regularization term. 

\subsection{Domain Translation based Cycle Feature Mapping} \label{method:mapping}

We assume that latent representations of cine imaging and mesh videos lie on separate manifolds, \(\mathcal{Z}_I\) and \(\mathcal{Z}_M\), with mapping functions \(G_M: \mathcal{Z}_I \to \mathcal{Z}_M\) and \(G_I: \mathcal{Z}_M \to \mathcal{Z}_I\). 
To estimate the corresponding 4D whole heart mesh for cine MRI, we need to perform a domain translation.
To achieve this, we employ generative adversarial networks to find the mapping between the two domains.
Specifically, the generator \(G_M\) learns to transform echo representations into realistic mesh video representations, while the discriminator \(D_M\) distinguishes real from generated samples, optimized via adversarial loss:  
\begin{equation}
    \mathcal{L}_M^{\text{adv}} = \mathbb{E}_{\phi_I} [\log D_M(G_M(\phi_I))] + \mathbb{E}_{\phi_M} [\log (1 - D_M(\phi_M))],
\end{equation}
\begin{equation}
    \mathcal{L}_I^{\text{adv}} = \mathbb{E}_{\phi_M} [\log D_I(G_I(\phi_M))] + \mathbb{E}_{\phi_I} [\log (1 - D_I(\phi_I))],
\end{equation}
where \( \phi_I \) and \( \phi_M \) are latent representations of cine and mesh videos, respectively.
To ensure consistency between the mapping functions \( G_I \) and \( G_M \), we employ a cycle-consistency loss:
\begin{equation}
    \mathcal{L}^{\text{cycle}} = \mathbb{E}_{\phi_I} \left[ \|G_I(G_M(\phi_I)) - \phi_I\|_1 \right] + \mathbb{E}_{\phi_M} \left[ \|G_M(G_I(\phi_M)) - \phi_M\|_1 \right],
\end{equation}
where \( \|\cdot\|_1 \) denotes the L1 norm.

To further improve the correspondences between cine and mesh videos, we introduce a pretrained EF prediction network \( N_{\text{EF}} \) to capture meaningful shape and dynamics information.
The ejection fraction (EF) is determined using the end-diastolic (ED) volume and end-systolic (ES) volume.
During training, we predict the EF from generated mesh representations \( G_M(\phi_I) \) and \( G_M(G_I(\phi_M)) \) using \( N_{\text{EF}} \). 
The EF loss is defined as:
\begin{equation}
    \mathcal{L}^{\text{EF}} = \mathbb{E}_{\phi_I} \left[ \|N_{\text{EF}}(G_M(\phi_I)) - \text{EF}_I\|_1 \right] + \mathbb{E}_{\phi_M} \left[ \|N_{\text{EF}}(G_M(G_E(\phi_M))) - \text{EF}_M\|_1 \right],
\end{equation}
where \( \text{EF}_I \) and \( \text{EF}_M \) are the ground-truth EF values from the cine and mesh data, respectively. 

\subsection{Personalized Inference of 4D Whole Heart Mesh} \label{method:inference}

Two domain-specific AEs can be independently trained, after which their weights are fixed. 
Subsequently, we can extract the compressed representations \(\phi_I\) and \(\phi_M\) from cine and mesh video data to train the domain shift modules. 
At the same time, the EF prediction network \(N_{\text{EF}}\) is trained on \(\mathcal{D}_M\).
Therefore, the mapping networks \(G_M\), \(G_E\), and discriminators \(D_E\), \(D_M\) are optimized based on the following objective:
\begin{equation}
    \min_{\lambda_{D_M}, \lambda_{D_E}} \max_{\lambda_{G_M}, \lambda_{G_E}} \beta_1 \mathcal{L}^{\text{adv}}_M + \beta_2 \mathcal{L}^{\text{adv}}_E + \beta_3 \mathcal{L}^{\text{cycle}} + \beta_4 \mathcal{L}^{\text{EF}},
\end{equation}
where \(\lambda_{D_M}\), \(\lambda_{D_E}\), \(\lambda_{G_M}\), and \(\lambda_{G_E}\) are the balancing parameters of the respective networks. 
Expectations in the loss functions are estimated via mini-batch sampling. 
Training stops when \(\mathcal{L}_{\text{EF}}\) on the validation sets stops decreasing.
During inference, the weights of all networks are fixed, and mesh videos are generated for each cine data using:
\begin{equation}
    \hat{\mathcal{M}} = \text{AE}^\text{Decoder}_M(G_M(\text{AE}^\text{Encoder}_I(I^{LAX}, I^{SAX}))).
\end{equation}
This process aims to generate personalized 4D meshes for each cine data.

\section{Experiments and Results}

\subsection{Materials}

\subsubsection{Data Acquisition and Pre-Processing.}

We collected 446 subjects with standard multi-view cardiac cine MRI. 
Specifically, three LAX slices and a SAX stack of 6–10 slices over one cardiac cycle are existed.
In total, 25-50 frames per cardiac cycle were obtained for each subject in the study population.
We only utilize the 4-chamber LAX and three representative slices from the SAX cine data.
All images were cropped into a unified size of 112 $\times$ 112 centering at the heart region, with a intensity normalization via Z-score. 
The dataset was randomly divided into 313 training and 133 test samples.
To train the mesh-specific AE, we generated 10,000 whole-heart surface mesh samples based on the SSM which embeds the morphological variation and dynamics observed in a cohort of 20 patients.

\subsubsection{Implementation.} 

The framework was implemented in TensorFlow, running on a computer with a 13th Gen Intel(R) Core(TM) i9-13980HX CPU and an NVIDIA GeForce RTX 4060 Laptop GPU.
We used the Adam optimizer to update the network parameters via stochastic gradient decent. 
The balancing parameters in Sec.~\ref{method:inference} are set as follows: \(\beta_1 = 1\), \(\beta_2 = 1\), \(\beta_3 = 10\), and \(\beta_4 = 10\).
The training of the model took about 65 hours, while the inference of one 4D heart from input cine images required about 18 min. 

\subsubsection{Gold Standard and Evaluation.}

Cine MRIs were manually segmented by a well-trained student using ITK-SNAP and checked by a senior expert. 
For the SAX view, LV, LV Myo, and RV have been labeled, while for the LAX view, LV, LV Myo, RV, LA, RA have been labeled.
These manual labels have been converted into contours, which are considered as ground truth in this work. 
For evaluation, we employed average surface distance (ASD) to assess the alignment between the predicted heart and the corresponding contours. 
Note that since the coordinate information was not learned in the generated mesh, we performed an initial alignment based on iterative closest point (ICP) registration algorithm between the contour and generated mesh for validation.

\subsection{Results}

\begin{table*} [t] \center
    \caption{              
    Summary of the quantitative evaluation results of 4D heart reconstruction in terms of average surface distance (mm).
     }
\label{tb:result:ASD}
{ \footnotesize 
\begin{tabular}{  l|l| l l l l l | l *{8}{@{\ \,} l }}
\hline
View & Phase  &  Myo  & LV & RV & LA & RA & Avg \\
\hline
\multirow{2}*{LAX} & ED    & $ 6.12 \pm 1.39 $ & $5.34 \pm 1.14$ & $10.7 \pm 2.29$ & $7.19 \pm 1.32 $ &  $6.08 \pm 1.39$ &  $7.08$ \\
& ES   & $7.66 \pm 1.83$ & $6.55 \pm 1.46$ & $9.28 \pm 0.90$ & $5.97 \pm 0.98$ & $5.38 \pm 1.07$  &  $6.97$ \\
\hline \hline
\multirow{2}*{LAX+SAX}  & ED    & $ 5.39 \pm 0.94 $ & $5.66 \pm 1.98$ & $11.3 \pm 2.74$ & $7.20 \pm 1.60 $ &  $5.69 \pm 1.24$ &  $7.06$ \\
& ES   & $6.41 \pm 1.52$ & $5.60 \pm 0.91$ & $9.72 \pm 1.55$ & $5.82 \pm 1.04$ & $5.80 \pm 1.21$  &  $6.67$ \\
\hline
\end{tabular} }\\
\end{table*}

\subsubsection{Effect of the Number of Cine Views.} 


To investigate the impact of the number of input views, we compared the performance of the proposed model using a single LAX view with that of a model incorporating both SAX and LAX views.
We randomly selected 20 subjects and manual segmented their LAX and SAX view at the ED and ES phases.
\Leireftb{tb:result:ASD} presents the average ASD values between reconstructed mesh and ground truth contours of the 20 subjects. 
One can see that in general combining multi-view cine MRI can perform slightly better than using single LAX view (ED: 7.06 mm vs. 7.08 mm and ES: 6.67 mm vs. 6.97 mm).
This is reasonable, as incorporating both views provided complementary spatial information for capturing complex cardiac structures and motion patterns across different perspectives. 
Multi-view fusion also mitigated errors or ambiguities from a single view, leading to more robust reconstructions. 
However, the improvement remained limited, as the model relies on learned priors and shape constraints, reducing sensitivity to additional views. 
Also, the weakly supervised nature of the model prevented explicit learning of anatomical correspondences, limiting the benefits of multi-view integration.

\begin{figure*}[t]\center
 \includegraphics[width=0.74\textwidth]{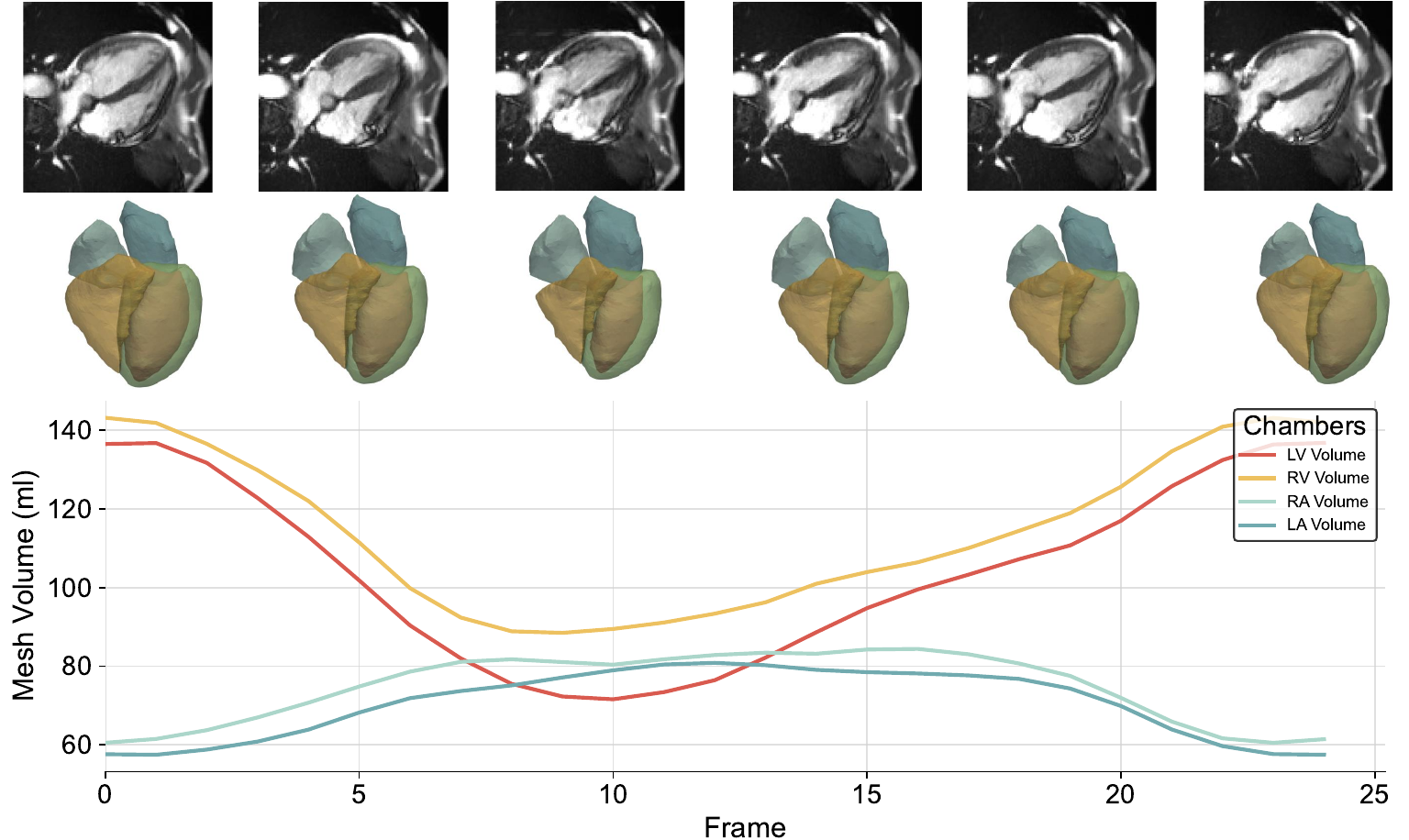}\\[-2ex]
   \caption{Illustration of cine MRI and predicted whole heart mesh with corresponding four chamber volume change over time.}
\label{fig:result:volume}
\end{figure*}

\subsubsection{Accuracy of the Reconstructed 4D Heart.}

Figure \ref{fig:result:volume} presents an example of the predicted heart shapes alongside the corresponding chamber volume changes over time. 
One can see that the reconstructed meshes accurately captured the cardiac systolic and diastolic phases observed in the cine MRI, demonstrating effective preservation of cardiac motion dynamics. 
Additionally, Fig. \ref{fig:result:EF} (a) shows the overlap visualization of predicted mesh and the ground truth in three randomly selected subjects.
One can see that the predicted chamber volume changed generally align with the ground truth, further validating the capability of the proposed model to adapt to both the shape and motion patterns of the heart. 
This consistency suggests that the learned representations effectively encode physiological deformations, enabling accurate and realistic 4D reconstructions.

\subsubsection{Capability of the Reconstructed 4D Heart for EF Estimation.} 

\begin{figure*}[t]\center
    \subfigure[] {\includegraphics[width=0.32\textwidth]{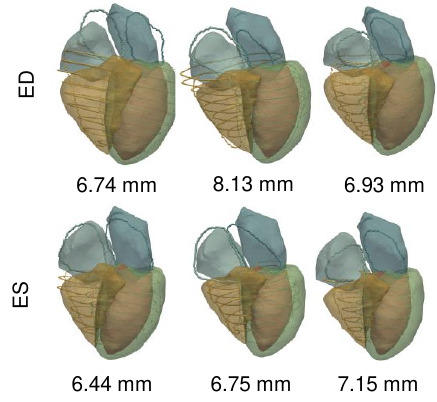}}
    \subfigure[] 
    {\includegraphics[width=0.3\textwidth]{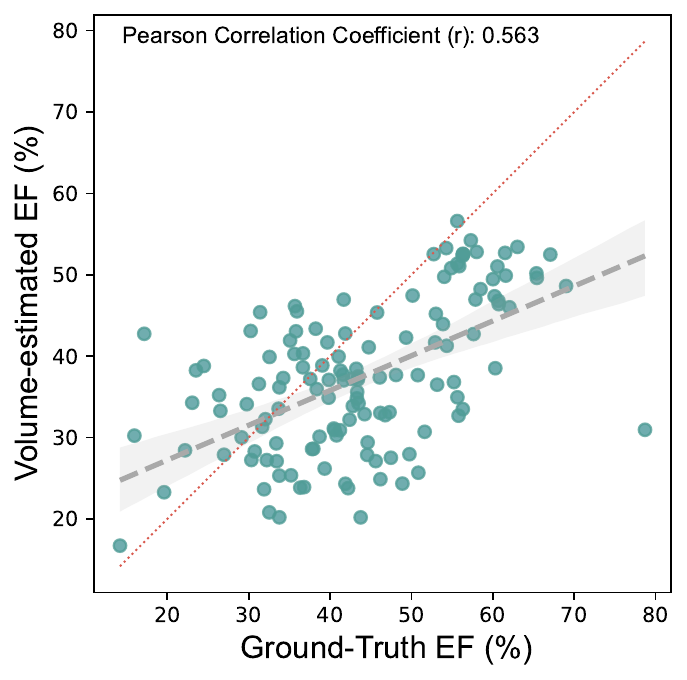}}
  \caption
    {(a) 3D visualization of the overlap between the sparse contours and reconstructed heart; (b) The scatter point and correlations between the ground truth and estimated EF.}
\label{fig:result:EF}
\end{figure*}

To further assess the accuracy of the reconstructed mesh, we compared EF values derived from manual segmentation of cine MRIs and predicted mesh, visualizing as a scatter plot for all test subjects (see Fig. \ref{fig:result:EF} (b)). 
To quantify this correlation, we performed linear regression and Pearson correlation analyses. 
The Pearson correlation coefficient of 0.563 indicated a moderate to strong correlation, suggesting that the reconstructed 4D heart model effectively captured volumetric changes during the cardiac cycle. 
However, some discrepancies may arise due to limited diversity of cardiac morphology and motion dynamics in the reconstruction process, which could restrict its ability to fully capture patient-specific variations. 
Further refinement, such as incorporating additional anatomical constraints, may enhance the accuracy of EF estimation.

\section{Conclusion}

In this work, we have proposed an end-to-end learning framework for automatic 4D whole heart mesh reconstruction by combining the LAX and SAX cine MRIs. 
The proposed algorithm has been applied to 446 subjects and obtained promising results compared manual delineation. 
The results have demonstrated the effectiveness of the proposed mapping scheme and showed the feasibility of employing 2D(+t) images for 3D(+t) reconstruction. 
Particularly, the proposed model does not rely on any paired imaging and mesh data nor the corresponding manual mask of cine MRIs.
Despite its effectiveness, this work has three main limitations.  
First, the input mesh distribution is derived from a small cohort of 20 subjects, limiting its ability to capture diverse cardiac shapes and motion patterns.  
Second, the reconstructed mesh lacks original image coordinates, which could be addressed using spatial constraints like cardiac landmarks or segmentation priors.  
Finally, further investigation is needed to validate the integration of the 4D mesh into cardiac simulations for efficient digital twin construction.

\bibliographystyle{splncs04}
\bibliography{A_refs}

\begin{thebibliography}{10}
\providecommand{\url}[1]{\texttt{#1}}
\providecommand{\urlprefix}{URL }
\providecommand{\doi}[1]{https://doi.org/#1}

\bibitem{journal/NC/arevalo2016}
Arevalo, H.J., Vadakkumpadan, F., Guallar, E., Jebb, A., Malamas, P., Wu, K.C.,
  Trayanova, N.A.: Arrhythmia risk stratification of patients after myocardial
  infarction using personalized heart models. Nature Communications
  \textbf{7}(1),  11437 (2016)

\bibitem{journal/PTRSA/banerjee2021}
Banerjee, A., Camps, J., Zacur, E., Andrews, C.M., Rudy, Y., Choudhury, R.P.,
  Rodriguez, B., Grau, V.: A completely automated pipeline for {3D}
  reconstruction of human heart from {2D} cine magnetic resonance slices.
  Philosophical Transactions of the Royal Society A  \textbf{379}(2212),
  20200257 (2021)

\bibitem{journal/MedIA/beetz2023}
Beetz, M., Banerjee, A., Ossenberg-Engels, J., Grau, V.: Multi-class point
  cloud completion networks for {3D} cardiac anatomy reconstruction from cine
  magnetic resonance images. Medical Image Analysis  \textbf{90},  102975
  (2023)

\bibitem{journal/BMM/bennati2023}
Bennati, L., Giambruno, V., Renzi, F., Di~Nicola, V., Maffeis, C., Puppini, G.,
  Luciani, G.B., Vergara, C.: Turbulent blood dynamics in the left heart in the
  presence of mitral regurgitation: a computational study based on multi-series
  cine-{MRI}. Biomechanics and Modeling in Mechanobiology  \textbf{22}(6),
  1829--1846 (2023)

\bibitem{conf/ISBI/biffi2019}
Biffi, C., Cerrolaza, J.J., Tarroni, G., de~Marvao, A., Cook, S.A., O’Regan,
  D.P., Rueckert, D.: {3D} high-resolution cardiac segmentation reconstruction
  from {2D} views using conditional variational autoencoders. In: nternational
  Symposium on Biomedical Imaging. pp. 1643--1646. IEEE (2019)

\bibitem{conf/MICCAI/chang2022}
Chang, Q., Yan, Z., Zhou, M., Liu, D., Sawalha, K., Ye, M., Zhangli, Q.,
  Kanski, M., Al’Aref, S., Axel, L., et~al.: Deeprecon: Joint {2D} cardiac
  segmentation and {3D} volume reconstruction via a structure-specific
  generative method. In: International Conference on Medical Image Computing
  and Computer-Assisted Intervention. pp. 567--577. Springer (2022)

\bibitem{journal/MedIA/chen2021}
Chen, X., Ravikumar, N., Xia, Y., Attar, R., Diaz-Pinto, A., Piechnik, S.K.,
  Neubauer, S., Petersen, S.E., Frangi, A.F.: Shape registration with learned
  deformations for {3D} shape reconstruction from sparse and incomplete point
  clouds. Medical Image Analysis  \textbf{74},  102228 (2021)

\bibitem{journal/arxiv/chen2024}
Chen, Y., Yang, J., Mercadier, D.S., Le, H., Fua, P.: Medtet: An online motion
  model for {4D} heart reconstruction. arXiv preprint arXiv:2412.02589  (2024)

\bibitem{journal/EHJ/corral2020}
Corral-Acero, J., et~al.: The ‘digital twin’to enable the vision of
  precision cardiology. European heart journal  \textbf{41}(48),  4556--4564
  (2020)

\bibitem{journal/MedIA/gillette2021}
Gillette, K., et~al.: A framework for the generation of digital twins of
  cardiac electrophysiology from clinical 12-leads {ECG}s. Medical Image
  Analysis  \textbf{71},  102080 (2021)

\bibitem{journal/CMPB/hu2023}
Hu, H., Pan, N., Frangi, A.F.: Fully automatic initialization and segmentation
  of left and right ventricles for large-scale cardiac {MRI} using a deeply
  supervised network and 3d-{ASM}. Computer Methods and Programs in Biomedicine
   \textbf{240},  107679 (2023)

\bibitem{journal/MedIA/kong2021}
Kong, F., Wilson, N., Shadden, S.: A deep-learning approach for direct
  whole-heart mesh reconstruction. Medical image analysis  \textbf{74},  102222
  (2021)

\bibitem{journal/MedIA/laumer2023}
Laumer, F., Amrani, M., Manduchi, L., Beuret, A., Rubi, L., Dubatovka, A.,
  Matter, C.M., Buhmann, J.M.: Weakly supervised inference of personalized
  heart meshes based on echocardiography videos. Medical image analysis
  \textbf{83},  102653 (2023)

\bibitem{journal/TMI/li2024}
Li, L., Camps, J., Wang, Z., Beetz, M., Banerjee, A., Rodriguez, B., Grau, V.:
  Towards enabling cardiac digital twins of myocardial infarction using deep
  computational models for inverse inference. IEEE Transactions on Medical
  Imaging  (2024)

\bibitem{journal/PTRS/niederer2020}
Niederer, S., et~al.: Creation and application of virtual patient cohorts of
  heart models. Philosophical Transactions of the Royal Society A
  \textbf{378}(2173),  20190558 (2020)

\bibitem{journal/MRM/odille2018}
Odille, F., Bustin, A., Liu, S., Chen, B., Vuissoz, P.A., Felblinger, J.,
  Bonnemains, L.: Isotropic {3D} cardiac cine {MRI} allows efficient sparse
  segmentation strategies based on {3D} surface reconstruction. Magnetic
  resonance in medicine  \textbf{79}(5),  2665--2675 (2018)

\bibitem{journal/TMI/ukwatta2015}
Ukwatta, E., Arevalo, H., Li, K., Yuan, J., Qiu, W., Malamas, P., Wu, K.C.,
  Trayanova, N.A., Vadakkumpadan, F.: Myocardial infarct segmentation from
  magnetic resonance images for personalized modeling of cardiac
  electrophysiology. IEEE Transactions on Medical Imaging  \textbf{35}(6),
  1408--1419 (2015)

\bibitem{journal/JI/villard2018}
Villard, B., Grau, V., Zacur, E.: Surface mesh reconstruction from cardiac
  {MRI} contours. Journal of Imaging  \textbf{4}(1), ~16 (2018)

\bibitem{conf/ICCV/ye2023}
Ye, M., Yang, D., Kanski, M., Axel, L., Metaxas, D.: Neural deformable models
  for {3D} bi-ventricular heart shape reconstruction and modeling from {2D}
  sparse cardiac magnetic resonance imaging. In: Proceedings of the IEEE/CVF
  International Conference on Computer Vision. pp. 14247--14256 (2023)

\bibitem{conf/ICCV/yuan2023}
Yuan, X., Liu, C., Wang, Y.: {4D} myocardium reconstruction with decoupled
  motion and shape model. In: Proceedings of the IEEE/CVF International
  Conference on Computer Vision. pp. 21252--21262 (2023)

\end{thebibliography}

\end{document}